\documentclass{article} 
\usepackage{arxiv_version}

\usepackage{microtype}
\usepackage{hyperref}
\usepackage{url}
\usepackage{booktabs}
\usepackage{amsmath}
\usepackage{amssymb}
\usepackage{authblk}
\usepackage{enumitem}
\usepackage{graphicx}
\usepackage{subcaption}
\usepackage{pythonhighlight}
\usepackage{multirow}

\title{AgentLite: A Lightweight Library for Building and Advancing Task-Oriented LLM Agent System}
 
\colmfinalcopy

\author{
\text{Zhiwei~Liu}\thanks{zhiweiliu@salesforce.com},
\text{Weiran~Yao},
\text{Jianguo~Zhang},
\text{Liangwei~Yang},
\text{Zuxin~Liu},
\text{Juntao~Tan}
\text{Prafulla K. Choubey},
\text{Tian Lan},
\text{Jason~Wu},
\text{Huan Wang},
\text{Shelby Heinecke}\\
\text{Caiming Xiong},
\text{Silvio Savarese}
}
\affil{Salesforce AI Research, USA}

\begin{document}

\maketitle

\begin{abstract}
The booming success of LLMs initiates rapid development in LLM agents. Though the foundation of an LLM agent is the generative model, it is critical to devise the optimal reasoning strategies and agent architectures.
Accordingly, LLM agent research advances from the simple chain-of-thought prompting to more complex ReAct and Reflection reasoning strategy; agent architecture also evolves from single agent generation to multi-agent conversation, as well as multi-LLM multi-agent group chat. 
However, with the existing intricate frameworks and libraries, creating and evaluating new reasoning strategies and agent architectures has become a complex challenge, which hinders research investigation into LLM agents.  
Thus, we open-source a new AI agent library, AgentLite, which simplifies this process by offering a lightweight, user-friendly platform for innovating LLM agent reasoning, architectures, and applications with ease. AgentLite is a task-oriented framework designed to enhance the ability of agents to break down tasks and facilitate the development of multi-agent systems.
Furthermore, we introduce multiple practical applications developed with AgentLite to demonstrate its convenience and flexibility. 
Get started now at: \url{https://github.com/SalesforceAIResearch/AgentLite}. 
\end{abstract}

\section{Introduction}
The AI agent emerges as a trending research topic along with the wide success of generative AI~\citep{openai2023gpt4,touvron2023llama}.
Among the outstanding innovations, the LLM agent augments an large language model~(LLM) with tool usage~\citep{patil2023gorilla,schick2023toolformer,qin2023toolllm} and environment interaction ability~\citep{liu2023agentbench,deng2023mind2web,zhou2023webarena}. 
As such, an agent is capable of collecting real-time feedback from environments to close the interaction loop~\citep{liu2023bolaa,yao2023react,xu2023rewoo},
which expands the task-solving capability of LLM from one-time generation to multi-step action execution~\citep{deng2023mindweb,zheng2023seeact,wu2023autogen}. 
Therefore, LLM agents exhibit promising performance in various LLM applications, such as BabyAGI~\citep{babyagi23}, AutoGPT~\citep{autogpt23}, agent world~\citep{Park2023GenerativeAgents} and etc~\citep{wu2023autogen,langchain23}.

Though the core of the LLM agent is the generative model for information understanding and action generation, the reasoning strategies and agent architectures are also rather crucial for enhancing performance.
Regarding the reasoning strategies, ReAct~\citep{yao2023react} exhibits higher execution correctness when activating one-step of \textit{Think} in agent action execution.
Also, enabling reflection~\citep{shinn2023reflexion,yao2023retroformer} reasoning enhances the performance of agents.
ReWOO~\citep{xu2023rewoo} decouples reasoning and observation in agent execution. 
Divergent Think~\citep{wang2023drdt} is also verified as one effective reasoning strategy for agents to explore more action spaces.

\begin{table}[!ht]
\centering
\small
\begin{tabular}{l|c|c|c|c|c}
Library        &  \textbf{AgentLite}               & AutoGen &  LangChain & Camel & CrewAI  \\ \hline
Task Decomposition & \checkmark       & -       & -       & \checkmark & \checkmark \\
Multi-Agent Orchestration      & \checkmark       & \checkmark       & -       & \checkmark & \checkmark \\
Extendable Reasoning Types    & \checkmark       & -      & \checkmark       & - & -  \\
Memory Module  & \checkmark    &  -   & \checkmark       &  \checkmark &  - \\ 
Prompter Module  & \checkmark &  -      & \checkmark       &  \checkmark  & - \\ 
\# (Core) Code Lines      & \textbf{959}       & 8,966       & 248,650       & 4,987 & 1,504  \\
\end{tabular}
\caption{
A comparison of features supported by AgentLite library vs existing libraries for agent building.
\checkmark denotes inherently supported in the library
while - represents not modulated or hard to refactor.}
\label{tab:lib_comparison}
\end{table}

Besides, the optimal agent architecture is yet not well established. 
Langchain~\citep{langchain23} evolves rapidly as one of the AI agent libraries, simplifying customized tool agent development.
BOLAA~\citep{liu2023bolaa} compares different agent architectures and proposes to use a multi-agent framework to decompose the complexity of tasks.
AutoGen~\citep{wu2023autogen} develops a novel multi-agent conversation mechanism and leverages code execution as actions.
$\alpha-$UMi~\citep{shen2024small} introduces a multi-LLM agent.
It fine-tunes multiple small LLMs in a multi-agent framework, which even outperforms GPT-4. 
Therefore,
more exploration~\citep{liu2023agentbench,zeng2023agenttuning} in regards to agent architecture designs are essential to advance the developments of LLM agents. 

In this sense, speeding up the implementation and validation of new reasoning types and agent architecture designs consolidates the fast development of LLM agents. 
As such,
a lightweight, easy-to-use library for efficiently prototyping LLM agent design is necessary.
Nevertheless, the existing libraries~\citep{langchain23,wu2023autogen,autogpt23,li2023camel}, though showing great progress, do not target at those fundamental designing of agents.
For example, Langchain, as one of the representative libraries for developing LLM applications, provides built-in interfaces to initialize different types of agents.
However, Langchain is over-complicated for researchers to implement new agent reasoning types and architectures.
Moreover, refactoring Langchain library for new research scenarios is rather challenging due to a lot of overhead in its agent creation.
Autogen, though achieving success in empowering LLM agent building, has hard-coded reasoning types in its agent interface, which requires a lot of effort for refactoring.
Also, its agent architecture is fixed as multi-agent conversation and code execution, which may not be suitable for new research scenarios and benchmarks.

To this end, we introduce a new AI agent building library, \textit{i.e.} AgentLite. 
AgentLite is a light research-oriented library for developing and implementing advanced LLM-based agent/multi-agent architectures.  
Building an agent with customized tools in AgentLite is easy and straightforward. AgentLite also enables the easy orchestration of multiple agents via a manager agent. 
The key features of AgentLite are as follows:
\begin{itemize}[leftmargin=*]

    \item \textbf{Lightweight Code Architecture}: AgentLite is characterized by its lean codebase, necessitating minimal external dependencies, thereby affording researchers unparalleled flexibility in customizing various agent components such as prompts, memory, action sequences, and Large Language Models (LLMs). AgentLite remains both accessible and adaptable to a wide range of research needs.

    \item \textbf{Task-Oriented Design Principle}: At the core of AgentLite is its focus on task-oriented agent development. 
    From the moment of initialization, each agent is endowed with a specific task, setting the stage for intricate task-based interactions.
    This foundational principle not only simplifies the design and deployment of individual agents but also serves as the basis for the construction of sophisticated multi-agent systems, facilitating seamless communication and collaboration among agents within the AgentLite ecosystem. 
    Therefore, AgentLite has great potential to develop a scalable framework for complex agent interactions.
 
    \item \textbf{Hierarchical Multi-Agent Orchestration}: AgentLite is setup upon a structured multi-agent systems. 
    This is achieved by assembling a team of agents, and each is governed by a manager agent responsible for orchestrating their operations.
    Such an arrangement endows the multi-agent system with a hierarchical structure, wherein the actions of team agents are coordinated to fulfill broader task objectives.
    
    \item \textbf{Versatile Implementation Potential}: AgentLite distinguishes its capacity for benchmarking evaluations and deployment in real-world scenarios. 
    AgentLite not only demonstrates its practical applicability in a variety of contexts but also underscores its utility in advancing the state-of-the-art in LLM agent applications.
\end{itemize}

A comparison of existing agent building library is presented in Table~\ref{tab:lib_comparison}, including AutoGen~\citep{wu2023autogen}, LangChain~\citep{langchain23}, Camel~\citep{li2023camel} and CrewAI\footnote{\url{https://github.com/joaomdmoura/crewAI/}}.
AgentLite achieves comprehensive abilities of LLM agents with less than 1k lines of code.

\section{AgentLite Framework}
\begin{figure}[ht]
    \centering
    \begin{subfigure}[b]{0.45\textwidth}
        \centering
        \includegraphics[height=0.9\textwidth]{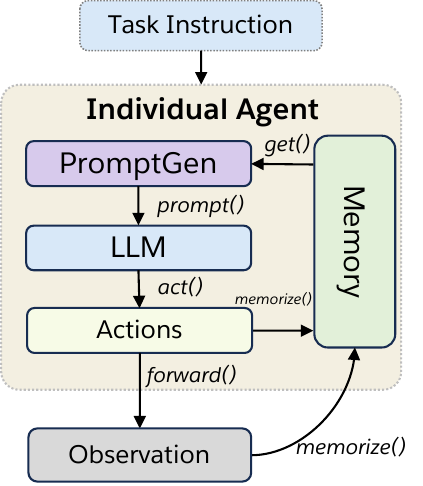}
        \caption{Individual Agent}
    \label{fig:individual}
    \end{subfigure}
    \begin{subfigure}[b]{0.45\textwidth}
        \centering
        \includegraphics[height=0.9\textwidth]{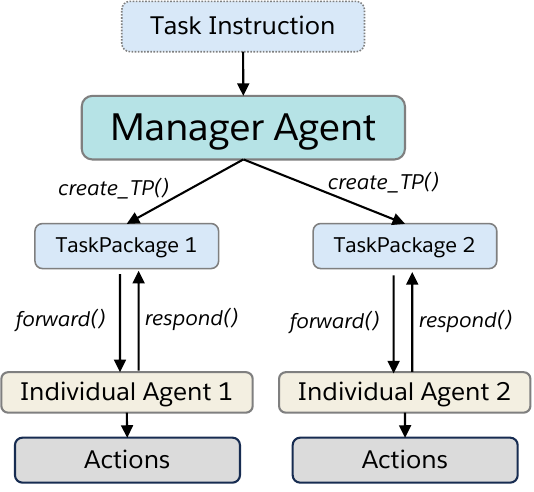}
        \caption{Manager Agent}
    \label{fig:manager_agent}
    \end{subfigure}
    \caption{The AgentLite framework for Individual Agent and Manager Agent. (a) Individual Agent is the base agent class in AgentLite.
    It constitutes four modules: \textit{i.e.} \textit{PromptGen}, \textit{LLM}, \textit{Actions} and \textit{Memory}. 
    The PromptGen composes a prompt for LLM from the memory of the agent. 
    LLM generates an executable action, which is forwarded to get observation.
    Both action and observation are memorized into the memory.
    (b) Manager Agent is a subclass of an individual agent with more methods in TaskPackage (TP) creation.
    A manager agent creates sub-tasks and corresponding TPs from the given task instruction.
    Those TPs are forwarded to associated individual agents sequentially. 
    }
    \label{fig:framework}
    \vspace{-1cm}
\end{figure}

We introduce the AgentLite framework by illustrating the general steps to implement customized Individual Agent and Manager Agent, our two cornerstones in the framework, for building a multi-agent system. The overview of our framework is illustrated in Figure~\ref{fig:framework}.

\subsection{Individual Agent}
Individual Agent is the base agent class in AgentLite. 
It is built upon four modules, i.e., PromptGen, Actions, LLM, and Memory, which are illustrated in Figure~\ref{fig:individual}.

\textbf{PromptGen} is a module to construct the prompt that will be sent to LLM to generate the action. 
The agent prompt is composed of multiple components, including agent role description, agent instructions, constraint for generation, agent actions, and examples. 
AgentLite supports default methods to combine those components.
Additionally, AgentLite allows developers to design customized prompts for action generation.

\textbf{Actions} are classes that an agent has and can automatically execute.
AgentLite provides a BaseAction as a wrapper class for developers to subclass their actions, code as follows:

\begin{minipage}{\linewidth}
\begin{python}
class BaseAction:
    def __init__(
        self, 
        action_name: str,  # the name of this action
        action_desc: str,  # the detailed explanation of this action
        params_doc: dict   # params documentation for __call__()
    )-> None:
        self.action_name = action_name
        self.action_desc = action_desc
        self.params_doc = params_doc

    def __call__(self, **kwargs) -> str:
        """
        implement the Action call 
        """
        raise NotImplementedError
\end{python}
\end{minipage}

\textit{action\_name}, \textit{action\_desc} and \textit{params\_doc} are the required properties for the agent to understand how to use this action. 
Specifically, the \textit{params\_doc} contains the key-value pairs to describe all the input parameters for the \textit{\_\_call\_\_()} method. 
Any customized agent actions should subclass this \textit{BaseAction} and overwrite the \textit{\_\_call\_\_()} method to return a string observation. 
Note that the \textit{action\_name} could be different from the actual class name.
In this way, developers could change it more flexibly for various scenarios. 
During execution, LLM generates both an action name and its input parameters, which are used to identify the executable actions and its function call input.
Different actions of agents also lead to different reasoning abilities.
For example, an agent with \textit{Think} action is equivalent to the ReAct reasoning strategy. 

\textbf{LLM} is based on API call method. AgentLite has a simple \textit{BaseLLM} wrapper class, which is called with an input string and returns another string generated from LLM.

\textbf{Memory} stores the action-observation chains of an agent. During execution, an agent gets its historical actions and corresponding observations, which are concatenated and input to PromptGen. All the generated actions and returned observations are immediately saved into memory.

Initializing an individual agent in AgentLite is then as simple as the following code snippet:

\begin{minipage}{\linewidth}
\begin{python}
from agentlite.agents import BaseAgent

name = "agent_name"
role = "describe the roles of this agent"
actions = [Action1, Action2]
agent = BaseAgent(name=name, role=role, actions=actions)
\end{python}
\end{minipage}

A typical working flow of an individual agent is as follows: firstly, the user gives task instructions to an agent. 
Then, the agent uses its PromptGen to construct the prompt, which retrieves the historical action-observation chain regarding this task if there is more than one step execution.
After getting the full prompt, LLM generates an action for the agent to execute. And the observation from this action is memorized into memory. 
Note that the BaseAgent has default \textit{Finish} action, which is an action to complete the current task and return a response. 
We also develop a default \textit{Think} action to enable the reasoning step. 

\subsection{Manager Agent}\label{sec:manager}
\textit{ManagerAgent} is another core agent class in AgentLite. 
It subclasses from \textit{BaseAgent} with more ability to communicate with other agents. 
The manager agent is designed to endow the multi-agent system development in AgentLite. 
We design this multi-agent system to be hierarchical.
Specifically, one manager agent is able to decompose a task into sub-tasks and send them to different individual agents. 
As such, each individual agent is interpreted as one type of action of this manager agent when adding to the team of this manager agent.
Therefore, this multi-agent could be designed with a deeper hierarchy if setting one manager agent to be a team member of another manager agent. 

Initializing a manager agent in AgentLite is simple and straightforward.
Developers directly pass in other agents as the \textit{TeamAgents} of this manager agent. 
For example, 

\begin{minipage}{\linewidth}
\begin{python}
from agentlite.agents import ManagerAgent

# assuming we have already designed three agents
# as a team for manager agents to control
team = [agent_1, agent_2] 
name = "manager_agent"
role = "controlling multiple agents to complete task"
manager = ManagerAgent(name=name,role=role,TeamAgents=team)
\end{python}
\end{minipage}

A typical working flow of a manager agent is presented in Figure~\ref{fig:manager_agent}.
The user sends a task instruction to a manager agent. Then, each action of the manager agent is to assign a task to one of its team agents to complete. 
In this step, a manager agent creates a \textit{TaskPackage}, forwards this \textit{TaskPackage} to this team agent and waits for the response from this team agent. 
\textit{TaskPackage} subclasses a common \textit{pydantic.BaseModel} in AgentLite, which has the following main properties:

\begin{minipage}{\linewidth}
\begin{python}\label{py:TP}
class TaskPackage(BaseModel):
    instruction: str # task instruction
    completion: str # completion status
    creator: str # agent name that creates this task
    timestamp: str 
    answer: str  # response for this task
    executor: str # agent name that executes this task
\end{python}
\end{minipage}

A Task Package (TP) serves as a communication basis between a managing agent and its subordinate team agents. The primary role of the managing agent is to concentrate on the generation and delegation of sub-tasks to these team agents. 
Upon receiving a sub-task, a designated agent is responsible for resolving this sub-task and subsequently communicates the outcome back to manager agent. 
This modular approach facilitates the assembly of a multi-agent system, where a developer can incorporate multiple specialized agents under the supervision of a managing agent to address complex tasks requiring diverse actions. 
Harnessing the collective capabilities of its components enhances its efficiency and brings a lot of potentials to a scalable multi-agent system. 
Note that the generation of TPs follows a sequential order, with the creation of a subsequent TP predicated on the feedback received from the execution of its predecessor.
This sequential process ensures that the system dynamically adapts and responds to the evolving requirements of the initial task.

\section{Agent Development}
In this section, we introduce how researchers could use AgentLite as code base to develop their new reasoning type and architecture.
\subsection{New Reasoning Type}
Adding a new reasoning type in AgentLite is similar to adding a new action into an agent, which is easy to implement.
For example, an agent with the reasoning type as direct \textit{Act} has actions for execution. 
In comparison, an agent with ReAct~\citep{yao2023react} reasoning is taking one \textit{Think} step before actual execution. 
In AgentLite implementation, we unify those agents \textbf{inner actions} like \textit{Think} and \textit{Plan}, and other executable actions like \textit{Search} as the same \textit{Action} module for the agent to call. 
A sample implementation of a \textit{Think} action in AgentLite is as follows\footnote{The codes are simplified for legibility.}:

\begin{minipage}{\linewidth}
\begin{python}
class ThinkAction(BaseAction):
    action_name = "Think"
    action_desc = "Conduct thinking..."
    params_doc = {
        INNER_ACT_KEY: "this is your thinking response.."
    }
    
    def __call__(self, **kwargs):
        return DEF_INNER_ACT_OBS
\end{python}
\end{minipage}

Enabling ReAct reasoning of agent is then simplified as adding this new \textit{ThinkAction} into the action of an agent as:

\begin{minipage}{\linewidth}
\begin{python}
agent.actions += [ThinkAction]
\end{python}
\end{minipage}

To be more specific, LLM uses the \textit{action\_name} and \textit{action\_desc} to understand this action.
Developers are also able to change them to different prompts. 
We formulate the actual reasoning process of LLM to generate the parameters of \textit{ThinkAction}. 
In this way, the reasoning process and the tool usage execution are unified.  
We could also develop \textit{Plan} and \textit{Reflect} actions analogously. 

\subsection{New Agent Architecture}\label{agent_architecture}
AgentLite also supports new Agent Architectures, which can be easily configured within lines of Python code. It provides sufficient flexibility for users to design agents. 
This section illustrates two types of new Agent Architectures with AgentLite.

\textbf{Copilot Agent.}
A Copilot Agent is able to receive human instructions and complete tasks accordingly, which is useful for a wide range of practical applications. 
It can be easily built by defining a \textit{HumanInstruction} action, and adding the new action to the agent.
The new action should inherit from the BaseAction class in AgentLite, and the code is given as follows:

\begin{minipage}{\linewidth}
\begin{python}
from agentlite.actions import BaseAction

class HumanInput(BaseAction):
    action_name = "HumanInput"
    action_desc = "Obtain instruction from human."
    params_doc = {"question": "Questions for human."}
    
    def __call__(self, question):
        instruction = input(question)
        
        return f"My instruction is: {instruction}:
\end{python}
\end{minipage}

By adding the newly-defined \textit{HumanInput} Action to the Agent, we can obtain a Copilot Agent that is able to acquire instruction from humans.
The code for building a Copilot Agent is as follows:

\begin{minipage}{\linewidth}
\begin{python}
from agentlite.agents import BaseAgent

name = "Copilot_Agent"
role = "Copilot agent that is able to receive human instructions."
actions = [HumanInput(), Action_1, Action_2, ...]
copilot_agent = BaseAgent(name=name, role=role, actions=actions)
\end{python}
\end{minipage}

\textbf{Copilot Multi-Agent.} 
AgentLite also supports an arbitrary number of agents for multi-agent orchestration. For example, a Copilot Multi-Agent is a multi-agent system with a human agent to collect instructions from human input. 
Designing this agent structure with AgentLite is realized via adding a \textit{HumanAgent} into a manager agent.
A simple \textit{HumanAgent}
subclasses from an abstract agent class in AgentLite, \textit{i.e.} \textit{ABCAgent}. 

\begin{minipage}{\linewidth}
\begin{python}
from agentlite.agents import ABCAgent

class HumanAgent(ABCAgent):
    name = "Human_Agent"
    role = "Collect instructions from human."

    def __call__(self, question):
        human_instruction = input(question)

        return f"My instruction is: {instruction}"
\end{python}
\end{minipage}

Then a copilot multi-agent is defined as a manager agent to control other agents with this human agent as a team. 
A sample code block is as follows:

\begin{minipage}{\linewidth}
\begin{python}
team = [HumanAgent(), agent_1, agent_2, ...]
name = "manager_agent"
role = "controlling a human agent, agent_1, agent_2, ..."
constraints = "should follow human instructions to complete task"
copilot_multiagent = ManagerAgent(name=name, role=role, constraints=constraints, TeamAgents=team)
\end{python}
\end{minipage}

\textbf{Multi-LLM Multi-Agent.}
AgentLite also supports different LLMs as backbones for different Agents. 
Before we initialize the agent, we should provide different LLM configs\footnote{Those LLMs should be running in the backend.}. 
Then, an agent with a specific LLM is defined as follows:

\begin{minipage}{\linewidth}
\begin{python}
from agentlite.llm import LLMConfig
# Define a specific LLM agent
llm_config_dict = {"llm_name": "LLM_1"}
llm_config = LLMConfig(llm_config_dict)
llm = get_llm_backend(llm_config)
agent_1 = BaseAgent(llm, name=name, role=role, actions=actions)
\end{python}
\end{minipage}

Assuming we have multiple agents with different LLMs, we finally build them as a team into a manager agent as follows:

\begin{minipage}{\linewidth}
\begin{python}
team = [agent_1, agent_2, agent_3, ...]
multi_LLM_multi_agent = ManagerAgent(team=team)
\end{python}
\end{minipage}

Note the LLM for this ManagerAgent is also configurable through this LLMConfig.


\section{Benchmark}
In this section, we implement specific agents with AgentLite to evaluate its performance on different agent benchmarks~\cite{liu2023bolaa,ma2024agentboard}.
\subsection{Retrieval-Augmented Question Answering}
The HotPotQA dataset \citep{yang2018hotpotqa} is a benchmark for evaluating the capabilities of question-answering (QA) systems in the context of multi-hop reasoning across multiple documents, which is adopted as a benchmark for evaluating LLM agent~\cite{liu2023bolaa}.
The performance of models on the HotPotQA dataset is quantitatively assessed using two primary metrics: \textbf{F1-Score} and \textbf{Accuracy}. 

In AgentLite, we devise the action space for this task as three main actions: \textit{WikipediaSearch}, \textit{Think}, and \textit{Finish}. 
The \textit{WikipediaSearch} action involves querying the Wikipedia dump to retrieve relevant articles or passages. 
The \textit{Think} action represents the model's internal processing to integrate the retrieved information, reason about it, and formulate a hypothesis toward the answer. 
Finally, the \textit{Finish} indicates the model's decision to conclude the search and reasoning process and output the final answer.
The results comparison on different models is reported in Table~\ref{tab:hotpotqa_results}.






\begin{table}[ht]
\centering
\caption{Performance of AgentLite on HotPotQA Dataset}
\label{tab:hotpotqa_results}
\resizebox{\textwidth}{!}{%
\begin{tabular}{@{}lcccccc@{}}
\toprule
\multirow{2}{*}{LLM} & \multicolumn{2}{c}{Easy} & \multicolumn{2}{c}{Medium} & \multicolumn{2}{c}{Hard} \\ \cmidrule(l){2-7} 
                       & F1-Score      & Accuracy     & F1-Score       & Accuracy      & F1-Score      & Accuracy     \\ \midrule
GPT-3.5-Turbo-16k-0613 & 0.410         & 0.35         & 0.330          & 0.25          & 0.283         & 0.20         \\
GPT-4-0613             & 0.611         & \textbf{0.47}         & 0.610          & 0.48          & \textbf{0.527}         & \textbf{0.38}         \\
GPT-4-32k-0613         & \textbf{0.625}         & 0.46         & \textbf{0.644}          & \textbf{0.54}          & 0.520         & 0.37         \\ 
xLAM-v0.1         & 0.532         & 0.45 & 0.547          & 0.46          & 0.455        & 0.36         \\ 

\bottomrule

\end{tabular}%
}
\end{table}

Our experiments on the HotPotQA dataset with models including GPT-3.5-Turbo-16k-0613, GPT-4-0613, GPT-4-32k-0613, and our xLAM-v0.1 model. 
xLAM-v0.1~\citep{zhang2024agentohana} is our advanced large action model (LAM) trained on agent action trajectories from multiple environments, which is built based on the Mixtral 8x7b Mix-of-Expert LLM~\citep{jiang2024mixtral}.
The evaluation are reported
across easy, medium, and hard difficulty levels, demonstrate a clear trend of performance improvement with the evolution of model architectures and capacities.
The GPT-4 variants outperform the GPT-3.5 model across all difficulty levels, with the 32k variant of GPT-4 showing superior performance in the medium difficulty set.
Our xLAM-v0.1 model also outperforms GPT-3.5, which indicates the necessity of fine-tuning an LLM on diverse action trajectories for improving agent performance.
This also demonstrates AgentLite is adaptable for arbitrary LLMs.


\subsection{Web Browser Decision Making}
We also implement an agent with AgentLite for Webshop environment.
WebShop~\citep{yao2022webshop} is an simulation online shopping website environment with 1.18M real-world products.
The environment also provides human instructions as shopping tasks.
Each instruction is associated with one ground-truth product, and contains attribute requirements.
Most existing agent evaluation adopts this environment for benchmarking~\citep{liu2023bolaa,liu2023agentbench,ma2024agentboard}.
We follow AgentBoard~\citep{ma2024agentboard} environment setup to design the webshop agent in AgentLite. 
Implementing a webshop agent with AgentLite is simple and straightforward.
We only need to wrap up both actions with the \textit{BaseAction} class and insert them into the \textit{BaseAgent} class.
The webshop environment also provides a final reward for shopping. 
We employ the same test tasks as in the AgentBoard, including a total of 251 instances. 
Each task is classified into one of two categories: easy (182 tasks) or hard (69 tasks).
In Table~\ref{tab:webshop_results}, we report the average rewards obtained for the easy, hard, and overall tasks when various models are applied. 
GPT-4-32k performs much better than other LLMs, especially on hard tasks.
The reason is that its longer context length enables more information understanding ability. 
Our xLAM model performs comparable with GPT-3.5 with AgentLite implementation.

\begin{table}[ht]
\centering
\caption{Performance of AgentLite on Webshop Environment}
\label{tab:webshop_results}
{%
\begin{tabular}{@{}lccc@{}}
\toprule
\multirow{2}{*}{LLM} & \multicolumn{3}{c}{avg. reward} \\
\cmidrule(l){2-4} 
& easy & hard & all \\
 \midrule
GPT-3.5-Turbo-16k-0613  & 0.528   &  0.506   & 0.522   \\
GPT-4-0613 & \textbf{0.676}        & 0.674         & 0.676 \\ 
GPT-4-32k-0613 & 0.675        &  \textbf{0.696}        & \textbf{0.681} \\
xLAM-v0.1 &    0.532     &   0.512       &  0.524 \\
\bottomrule

\end{tabular}%
}
\end{table}

\section{Applications}
In this section, we present various applications built via AgentLite to show the simplified design principle and innovative application capacity. These examples showcase varied characteristics including Practical Application (A3, A4), Human-in-the-Loop (A1, A2), Multi-modality (A1, A2), Innovative Potentional (A5).

\begin{figure*}
    \centering
    \includegraphics[width=0.85\linewidth]{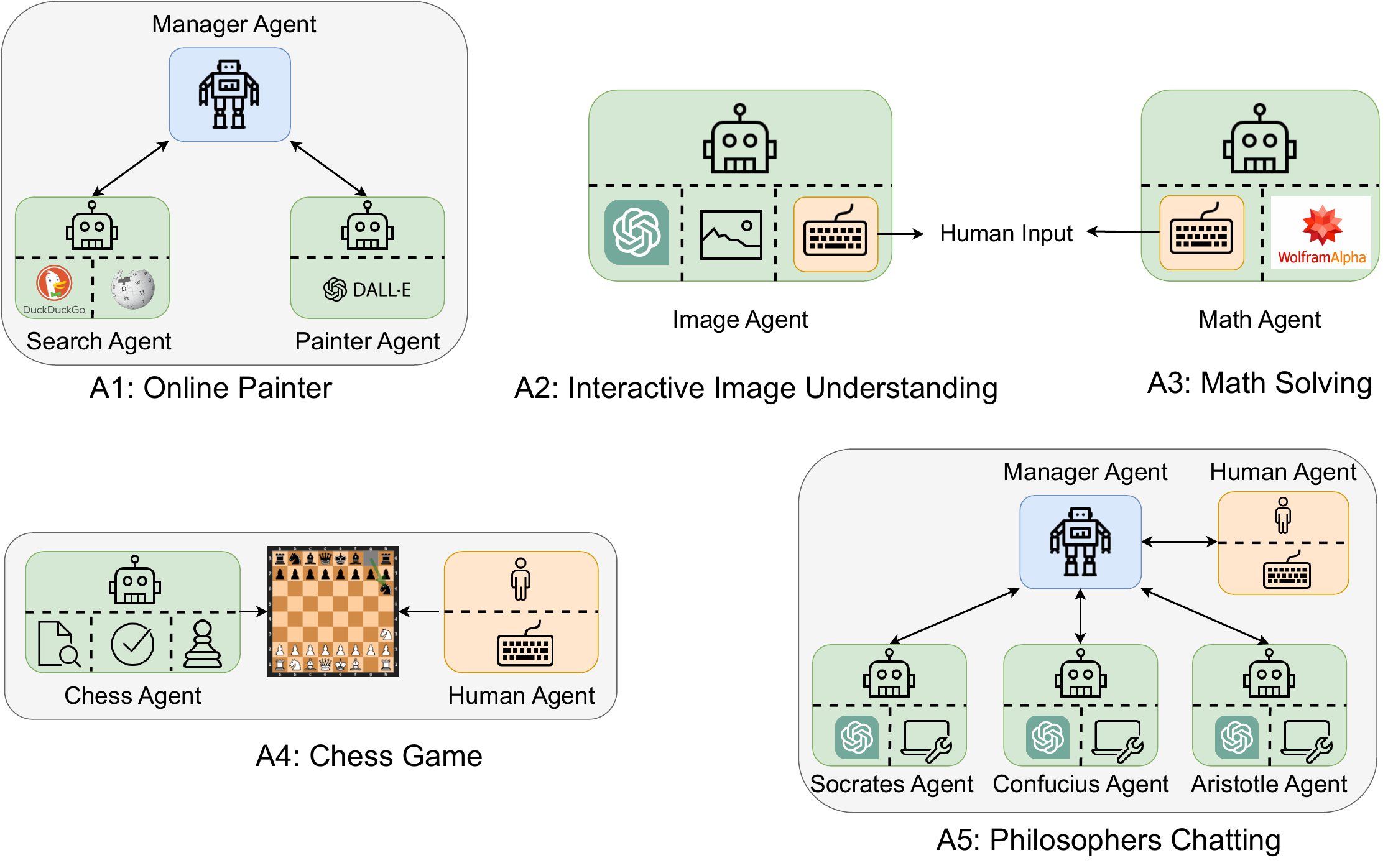}
    \caption{Application illustrations of AgentLite.}
    \label{fig:application}
\end{figure*}

\subsection{A1: Online Painter}

In this subsection, we illustrate the Online Painter application as shown in Fig.~\ref{fig:application}(A1). Given any object, Online Painter is able to search relevant visual features online first and then paint an image of the object based on the detailed features. 

The Online Painter application is completed by three agents (Manager Agent, Search Agent, Painter Agent). 
For example, the Painter Agent in AgentLite is initialized as follows:

\begin{minipage}{\linewidth}
\begin{python}
name = "Painter_agent"
role = "You are a painter and can draw a picture with Paint action."
actions = [Paint()]
PaintAgent = BaseAgent(name=name, role=role, actions=actions)
\end{python}
\end{minipage}

Search Agent can be defined analogously. 
Then, we could build one manager agent to control them as follows:

\begin{minipage}{\linewidth}
\begin{python}
manager = ManagerAgent(TeamAgents=[PaintAgent, SearchAgent])
\end{python}
\end{minipage}

The three agents all have the \textit{Think} action for reasoning and \textit{Finish} action to finish the task. 
Manager Agent receives tasks directly with the Task Package defined in Sec.~\ref{sec:manager} and dismantles the task to different steps for its team members consisting of Search Agent and Painter Agent.
The Search Agent have two actions to use for collecting information. 
    \textit{DuchSearch} collect online information via DuckDuckGo\footnote{\url{https://duckduckgo.com/}} API based on the query.
    And \textit{WikipediaSearch} is able to find relevant information on Wikipedia\footnote{\url{https://en.wikipedia.org/wiki/Main_Page/}}.
        These two actions endow the Search Agent with the ability to obtain information about the painted object.
(3) The Painter Agent have one DALLE\footnote{\url{https://openai.com/dall-e-2}} action to paint the picture given the description.

\subsection{A2: Interactive Image Understanding}
Interactive Image Understanding is a multi-modality application with human-in-the-loop instructions. Given an image, the application can provide answers to the questions from the human based on the image. Humans can give questions for several rounds to the application with natural languages such as \textit{"What is in the image?"}, \textit{"What is the color of the bridge?"}, etc. The application can also be terminated with ending instructions from humans such as \textit{"I good for today, have a great weekend."}, as real-life conversations.

The application is a copilot agent as described in Sec~\ref{agent_architecture}. The \textit{ImageDisplay} action can show the image to a human user and \textit{ImageQuery} action is able to answer questions based on the image. The \textit{ImageQuery} action is backended by the gpt-4-vision-preview api~\footnote{\url{https://platform.openai.com/docs/guides/vision}}. Empowered by the \textit{HumanInput} action, the Image Agent is able to acquire and follow instructions from humans. The code for building the Image Agent is as follows:

\begin{minipage}{\linewidth}
\begin{python}
from agentlite.agents import BaseAgent
name = "Image_agent"
role = "You are image agent to view and answer questions from image."
actions = [HumanInput(), ImageQuery(), ImageDisplay()]
ImageAgent = BaseAgent(name=name, role=role, actions=actions)
\end{python}
\end{minipage}

\subsection{A3: Math Problem Solving}
Math Problem Solving is an application built simply with a Math Agent via AgentLite. 
The Math Agent receives and solves a math problem from the human input. 
Human inputs can be either direct math equations as \textit{"75*34+12="} or descriptions of the problem as \textit{"What is the result of 75 multiplied by 34 and then plus 12?"}.
We use AgentLite to develop a Math Copilot Agent with \textit{HumanInput} and \textit{WolframAlphaSolver} actions. 
The Math Agent acquires math questions from humans with \textit{HumanInput} action, and then solve the problem with \textit{WolframAlphaSolver} action.
It is integrated with the \textit{WolframAlpha} API~\footnote{\url{https://www.wolframalpha.com/}}, which is able to solve a wide range of math problems from basic operations to equation solving and calculus.
The code for building the Math Problem Solving Copilo Agent is as follows:

\begin{minipage}{\linewidth}
\begin{python}
from agentlite.agents import BaseAgent
name = "MathAgent"
role = "You can answer math questions by WolframAlphaSolver action."
actions = [HumanInput(), WolframAlphaSolver()]
MathAgent = BaseAgent(name=name, role=role, actions=actions)
\end{python}
\end{minipage}

This is an example to show how to build applications directly with the individual agent. 
In AgentLite, the Manager Agent is not required for development.
Instead, it depends on the application's complexity and scale. 
For the simple Math Problem Solving application, an individual agent is sufficient. 
But for more complex scenarios involving multiple agents to work together as A1, A5, it is better to have a Manager Agent to organize the working flow of individual agents.

\subsection{A4: Chess Game}
Chess Game is a practical application to enable playing the Chess game against Large Language models. It takes turns to ask the Chess Agent and Human to move a step on the Chessboard. As the workflow can be easily managed, we do not need to build a Manager Agent here. It involves one Human Agent collecting input from the keyboard and one Chess Agent that is able to view the board, make analysis, and move the chess.

\begin{minipage}{\linewidth}
\begin{python}
from agentlite.agents import BaseAgent
name = "Chess_Player"
role = "You are a chess player to make move on the chess board."
actions = [BoardMove(), BoardView(), LegalMoves()]
Chess_Agent = BaseAgent(name=name, role=role, actions=actions)
\end{python}
\end{minipage}

Human Agent directly collects the human input from the keyboard and makes the corresponding move on the board. Chess Agent has $5$ actions to play chess with human. \textit{Think} and \textit{Finish} action are internally supported for reasoning and finish. 
Besides, $3$ more actions are further added to the Chess Agent. 
\textit{BoardView} action obtains current board's Forsyth-Edwards Notation (FEN), which is a standard notation to describe the positions of chess, and display the current board. 
\textit{LegalMoves} action obtains current available moves of the board.
Chess Agent finds available moves with this action.
\textit{BoardMove} action moves the chess on the board. 
To be noted that, agent can be on both side of the Chess Board, which further supports Agent-Agent Chess game.

\subsection{A5: Philosophers Chatting}
Philosophers Chatting is an innovative application to use the Large Language Model to mimic great Philosopher's thoughts including Socrates, Confucius, and Aristotle. The whole application is a copilot multi-agent as described in Sec.~\ref{agent_architecture}, which is a Manager Agent with its team members as Human Agent, Socrates Agent, Confucius Agent, and Aristotle Agent.
Each time Human can give one philosophy question to the Manager Agent such as \textit{"What should we pursue during our life?"}. After the Manager Agent receives the question, it asks the Socrates/Confucius/Aristotle agents for their opinion one by one and summarizes the results to answer the question.

To accomplish the application, three kinds of agents are involved. 
Human Agent directly provide philosophy questions for the Manager Agent.
Philosopher Agents (Socrates, Confucius, Aristotle) can represent different great philosophers in history and answer the question grounded by their literary work. For example, we could define \textit{Socrates\_Agent} as follows:

\begin{minipage}{\linewidth}
\begin{python}
from agentlite.agents import BaseAgent
name = "Socrates"
role = "You are Socrates. You are very familiar with Socrates's Book 
and Thought. Tell your opinion on behalf of Socrates"
Socrates_Agent = BaseAgent(name=name, role=role)
\end{python}
\end{minipage}


The Manager Agent is instantiated through the integration of four distinct agents into a team. 
Manager Agent orchestrates discussion flow, which encompasses identification of relevant questions, forwarding of these inquiries to appropriate agent, and summarization of the discussion.
This structured approach facilitates an effective exchange of information. 

\section{Conclusion and Future Work}
AgentLite is a lightweight codebase for researchers to develop customized LLM agent systems: \textbf{1)} One can easily use it for reproducing the benchmark results and prototyping various applications; \textbf{2)} AgentLite enables effortless integration and evaluation of new reasoning strategies and agent architectures. 
We develop a series of applications with AgentLite to showcase the superiority.
In the future, we will introduce additional communication methods among agents and offer a wider range of reasoning types in the agent class. 
\newpage
\bibliography{reference}
\bibliographystyle{colm2024_conference}


\end{document}